\documentstyle[preprint,aps]{revtex}
\begin{document}
\title{{\bf Generator Coordinate Calculations for
Breathing-Mode Giant Monopole Resonance in the Relativistic
Mean-Field Theory}\footnote{Dedicated to Prof. Dr. Klaus
Dietrich on the occasion of his 60th birthday}}
\author{M.V. STOITSOV$^{a,}$\cite{AAA},
        P. RING$^a$ and M.M. SHARMA$^b$}
\address{$^a$Physik Department, Technische Universit\"at
M\"unchen, D-85748 Garching, Germany}
\address{$^b$Max Planck Institut f\"ur Astrophysik,
Karl-Schwarzschild-Strasse 1, D-85740 Garching, Germany}

\maketitle

\begin{abstract}
\baselineskip=20pt
The breathing-mode giant monopole resonance (GMR) is
studied within the framework of the relativistic mean-field
theory using the Generator Coordinate Method (GCM). The
constrained incompressibility and the excitation energy of
isoscalar giant monopole states are obtained for finite
nuclei with various sets of Lagrangian parameters.  A
comparison is made with the results of nonrelativistic
constrained Skyrme Hartree-Fock calculations and with those
from Skyrme RPA calculations. In the RMF theory the GCM
calculations give a transition density for the breathing
mode, which resembles much that obtained from the Skyrme
HF+RPA approach and also that from the scaling mode of the
GMR.  From the systematic study of the breathing-mode as a
function of the incompressibility in GCM, it is shown that
the GCM succeeds in describing the GMR energies in nuclei
and that the empirical breathing-mode energies of heavy
nuclei can be reproduced by forces with an
incompressibility close to $K = 300$ MeV in the RMF theory.
\end{abstract}

\narrowtext

\section{INTRODUCTION}
\label{intro}

The nuclear matter incompressibility signifies an important
and cardinal point on the equation of state (EOS). The
behaviour of the nuclear matter at the saturation point is
relevant not only to the property of finite nuclei, but
also to astrophysical phenomena such as supernovae
explosion and neutron stars. The breathing-mode giant
monopole resonance (GMR), whereby nuclei undergo radial
density oscillations, provides a source of extracting the
dynamical behaviour, i.e., the compression properties of
nuclei and nuclear matter\cite{sha90}.  In addition to the
GMR excitation mode which represents a small-amplitude
collective motion, the intermediate energy heavy-ion
collisions\cite{stoe86}, on the other hand, strive to map
out the EOS of the nuclear matter for densities higher than
the saturation density. This is also expected to constrain
the incompressibility at the saturation point. However,
owing to the complex interplay of many degrees of freedom
in the heavy-ion collision, it has not yet been possible to
gain much insight into the behaviour of the EOS.  For
properties around the saturation point, the GMR remains an
important object of investigations.

The GMR has been measured almost all over the periodic
table\cite{bue81}.  Some time ago, the GMR energy was
obtained\cite{sha88} with a considerable precision in a set
of medium heavy Sn and Sm nuclei. Attempts were made to
extract the nuclear matter incompressibility from such
precision measurements. An earlier analysis based upon a
leptodermous expansion of finite nuclear incompressibility
into various finite-size components, led to the nuclear
matter incompressibility of $300 \pm 25$ MeV\cite{sha89}.
This analysis, which took into account the correlation of
the Coulomb term involving the third derivative of the EOS,
was based upon the systematics from the density-dependent
Skyrme interactions.  In a real case, however, Skyrme
forces might not be reliable for this purpose.
Circumventing this constraint based upon the Skyrme
interactions, it was found that error bars on the nuclear
matter incompressibility increased by more than 50\% and
the value itself was obtained at slightly higher than 300
MeV\cite{sha91}. More recently an analysis of experimental
data including deformed nuclei and of data from many
laboratories was attempted\cite{shlo93}.  However, this
analysis, which comprises data of various origins, was not
conclusive on the extraction of the nuclear matter
incompressibility. A detailed and critical analysis of
empirical breathing-mode GMR data is in progress.

Theoretically, the incompressibility has been obtained
using the density-dependent interactions\cite{bla80}. The
deductions base themselves upon an interpolation between
various Skyrme and Gogny forces for the GMR energies
obtained from self-consistent HF+RPA calculations. These
calculations were a major effort with a view to explaining
the breathing-mode energies in finite nuclei in a
microscopic approach. This approach, however, succeeded in
reproducing the GMR energy of only $^{208}$Pb within the
interpolation scheme. The GMR energies of $^{90}$Zr were
overestimated by 1-2 MeV. This fact has been corroborated
by the calculations within the RPA sum-rule approach using
various Skyrme interactions\cite{sha89}. The calculations
indeed reproduce the GMR energy of $^{208}$Pb using Skyrme
force SkM$^*$. The GMR energies of medium-heavy nuclei such
as $^{90}$Zr, Sn and Sm isotopes could not, however, be
reproduced within the Skyrme forces. The Skyrme interaction
SkM$^*$ has been used extensively to calculate the
properties of giant resonances\cite{gle90}. It reproduces
the empirical excitation energies of giant quadrupole
resonance (GQR) very well. The appropriate effective mass
of this force helps to achieve the required GQR energies.
The force SkM$^*$, however, reproduces the GMR energies of
only $^{208}$Pb well. This is due to a simple relationship
of the surface incompressibility to the bulk
incompressibility for the Skyrme type of forces, that for a
given force the surface incompressibility has about the
same value as the bulk incompressibility\cite{tre81}. This
relationship has essentially been at the root of the
problems in describing adequately the mass dependence of
the GMR energies in the Skyrme ansatz.

Relativistic mean-field (RMF) theory\cite{ser86} has in the
last years been found to be especially appealing in
describing the ground-state properties of nuclei at and far
away from the stability line\cite{gam90,snr93}.  The
long-standing problem of the kink in isotope shifts in Pb
nuclei, which could not be described with the Skyrme forces
including all possible correlations, has been successfully
solved in the RMF theory\cite{slr93}. The theory has
subsequently also been able to provide a good description
of the binding energies and deformations of nuclei close to
neutron drip-line\cite{slr94}.  Shell effects arising from
the Dirac structure of the spin-orbit interaction in the
RMF theory manifest in the behaviour of the binding
energies.  The strong shell-effects arising from the RMF
theory are corroborated by the finite-range droplet model
(FRDM)\cite{mol94} and are in contrast with those from the
Skyrme theory\cite{hae89,dob94}.  Thus, the RMF theory has
achieved a considerable success in describing many aspects
of the ground-state properties of nuclei.

The dynamical aspects within the RMF theory have remained
largely unexplored. A first attempt was made to obtain the
breathing-mode energies and incompressibilities within the
RMF theory using the linear Walecka model in constrained
calculations\cite{mar89}. Such calculations were further
extended to light nuclei and anharmonicities in the
breathing-mode oscillations were indicated\cite{malf91}.
The relationship of the GMR energies to the
incompressibility of nuclear matter is, however, not yet
known for the RMF theory. On the contrary, in the Skyrme
approach, the relationship between the GMR energies and the
incompressibility has been studied extensively (see e.g.,
refs.\cite{bla80,tre81,sha89,sha90,gle90}) and has been
found to be straightforward. An exercise to understand this
relationship in the RMF theory has recently been
undertaken\cite{srs94} employing  relativistic constrained
calculations within the mean field. Another approach which
has received considerable attention as a useful tool to
study properties of excited stated in nuclei is the
generator coordinate method (GCM)\cite{hil53}. It has been
applied amongst others also for the breathing
mode\cite{fer56,flo76,kre76}. This has been attempted in
the non-relativistic theories with a view to taking into
account the relevant correlations in the nuclei. In this
paper, we investigate the GCM for the first time in the RMF
theory and focus upon the structure and properties of the
breathing-mode GMR using the method of generator
coordinates. A comparison of the properties of the GMR will
be made with those from the Skyrme ansatz.

The paper is organized in the following way: In Section II
we provide the theoretical framework of the RMF theory. The
details on the Generator Coordinate Method in the RMF
theory are presented in Section III. The problem of the
breathing mode GMR is discussed in Section IV. In Section V
we discuss the results obtained in this framework. The last
section contains a summary and conclusions.

\section{Relativistic Mean-Field Theory}
\label{rmf-theory}

We start from Relativistic Mean Field theory\cite{ser86}
which treats the nucleons as Dirac spinors $\psi$
interacting by the exchange of several mesons: scalar
$\sigma$-meson that produces a strong attraction, isoscalar
vector $\omega$-meson that causes a strong  repulsion,
isovector $\rho$-meson required to generate the required
isospin asymmetry and photon that produces the
electromagnetic interaction. The model Lagrangian density
is:
\begin{equation}
\begin{array}{ll}
{\cal L}& =  \bar\psi \{ i\gamma_\mu\partial^\mu-M\}\psi
   +{1\over 2}\partial^\mu\sigma\partial_\mu\sigma
   -U(\sigma)-g_\sigma\bar\psi\sigma\psi \\
  & -{1\over 4}\Omega^{\mu\nu}\Omega_{\mu\nu}
   +{1\over 2}m^2_\omega \omega_\mu\omega^\mu
   -g_\omega\bar\psi\gamma_\mu\omega^\mu\psi  \\
  & -{1\over 4}\vec {R}^{\mu\nu}\vec {R}_{\mu\nu}
   +{1\over 2}m^2_\rho \vec {\rho}_\mu\vec {\rho}^\mu
   -g_\rho\bar\psi\gamma^\mu\vec{\tau}\psi\vec{\rho}_\mu  \\
  & -{1\over 4} F^{\mu\nu} F_{\mu\nu}
   -e\bar\psi\gamma^\mu{{(1-\tau_3)} \over 2} \psi A_\mu,
\label{lagrangian}
\end{array}
\end{equation}
where $U(\sigma)$ is the non-linear scalar self-interaction
with the cubic and quartic terms required for appropriate
surface properties\cite{bog77}:
\begin{equation}
U(\sigma)={1\over 2}m^2_\sigma\sigma^2+
{1\over 3}g_2\sigma^3+{1\over 4}g_3\sigma^4.
\label{nlpot}
\end{equation}
M, $m_\sigma$, $m_\omega$ and $m_\rho$ are the nucleon, the
$\sigma$-, the $\omega$-, and the $\rho$-meson masses,
respectively, and $g_\sigma$, $g_\omega$, $g_\rho$ and
$e^2$/$4\pi$=1/137 are the coupling constants for the
$\sigma$-,  the $\omega$-, the $\rho$-mesons and for the
photon. The field tensors for the vector mesons are:
\begin{equation}
\Omega^{\mu\nu}=\partial^\mu\omega^\nu-\partial^\nu\omega^\mu,
\end{equation}
\begin{equation}
\vec{R}^{\mu\nu}=\partial^\mu\vec{\rho}^\nu-\partial^\nu\vec{\rho}^\mu
- g_\rho(\vec {\rho}^\mu \times \vec{\rho}^\nu)
\end{equation}
and  for the electromagnetic field
\begin{equation}
{F}^{\mu\nu}=\partial^\mu{A}^\nu-\partial^\nu{A}^\mu.
\end{equation}
The associated Hamiltonian  operator $\hat H$ is then
obtained using the well known canonical quantization
procedure based on the anti-commutator (for the fermions)
and the commutator (for the mesons)
relations\cite{kuch91,stoit94}.

Within the relativistic mean-field (RMF) approximation the
A independent nucleons with single-particle spinors
$\psi_i,\; (i=1,2,...,A)$, are  assumed to form a
single Slater determinant $\Phi$ and to move independently
in the meson fields. In the particular case of spherical
nuclei, symmetries simplify the calculations considerably
and only the time-like components $\omega^0(r)$, $\rho^0(r)$
and $A^0(r)$ of the $\omega$-, the $\rho$- and the
electromagnetic fields survive. When describing groundstate
properties of nuclei, one looks for static field solutions
$\phi(r)=\sigma(r),\; \omega^0(r),\; \rho^0(r)$ and
$A^0(r)$ that satisfy the Klein-Gordon equation
\begin{equation}
     \left( -{{\partial^2}\over {\partial r^2}}
     -{2\over r}{\partial \over {\partial r}} + m^2_\phi \right)
                        \phi(r) = s_\phi(r),
\label{kg}
\end{equation}
where $m_\phi$ are the meson masses for $\phi=\sigma,\;
\omega, \; \rho$ and $m_\phi$ is zero  for the photon. The
source terms
\begin{equation}
    s_\phi(r) = \left\{ \begin{array}{ll}
-g_\sigma\rho_s(r)-g_2\sigma^2(r)-g_3\sigma^3(r)
&\mbox{for the$\;\sigma$-field}\\
g_\omega \rho_v(r)&\mbox{for the$\;\omega$-field}\\
 g_\rho \rho_3(r)&\mbox{for the$\;\rho$-field  }\\
 e\rho_p(r)&\mbox{for the Coulomb  field,}
\end{array}
\right.
\label{sources}
\end{equation}
depend on the spherical densities
\begin{equation}
\begin{array}{ll}
\rho_s(r)  &=\displaystyle{\sum_{i=1}^A}\;\bar\psi_i({\bf r}) \psi_i({\bf r})\\
\rho_v(r)  &=\displaystyle{\sum_{i=1}^A}\;\psi_i^\dagger({\bf r}) \psi_i({\bf
r})\\
\rho_3(r)  &=\displaystyle{\sum_{i=1}^A}\;
\psi_i^\dagger({\bf r})\tau_3\psi_i({\bf r})\\
\rho_p(r)  &=\displaystyle{\sum_{i=1}^A}\;\psi_i^\dagger({\bf r})
{{(1-\tau_3)}\over 2} \psi_i({\bf r}),
\end{array}
\label{densities}
\end{equation}
where, in the no-sea approximation, the summation runs over
all occupied states in the Slater determinant $\Phi$. The
solution of  Eq. (\ref{kg}) can, in principle, be expressed
in terms of Green's functions, i.e.,

\begin{equation}
\phi(r)=\displaystyle{\int_0^\infty}G_\phi(r,r')s_\phi(r')r'^2 dr',
\label{mesons}
\end{equation}
where, for the massive fields,
\begin{equation}
G_\phi(r,r')={1\over {2m_\phi}}{1\over
{rr'}}\left(e^{-m_\phi\mid r-r'\mid } - e^{-m_\phi\mid
r+r'\mid }\right),
\label{green-mes}
\end{equation}
and for the Coulomb field,

\begin{equation}
G_\phi(r,r')= \left\{ \begin{array}{ll}
                         1/r  & \mbox{for $r>r'$} \\
                         1/r' & \mbox{for $r<r'$. }
                        \end{array}
                \right.
\label{green-cou}
\end{equation}
The total ground-state energy of spherical nuclei can be
expressed, in the center-of-mass frame, as a functional of
the baryon spinors $\{\psi_i\}$
\begin{equation}
E_{RMF}[\psi_i]~\equiv~<\Phi|\hat {H}|\Phi>
\label{energy-functional}
\end{equation}
where the Hamiltonian density
\begin{equation}
\begin{array}{ll}
 {\cal H }_{RMF}(r)~=~\tau(r) &+ M \rho_s(r)
 + {1\over 2} g_\sigma\rho_s(r)\sigma(r)
 + {1\over 2}\{ {2\over 3} g_2\sigma^3(r)
                + {1\over 2} g_3\sigma^4(r) \} \\
 &+ {1\over 2} g_\omega\rho_v(r)\omega^0(r)
  + {1\over 2} g_\rho\rho_3(r)\rho^0(r)
  + {1\over 2} e\rho_p(r)A^0(r)
\end{array}
\label{energy-density}
\end{equation}
depends only on the baryon field since the 'kinetic' energy
density
\begin{equation}
\tau(r)~\equiv~\displaystyle{\sum_{i=1}^A}\;
\psi_i^\dagger({\bf r}) \{ -i{\mbox{\boldmath $\alpha$}}
{\mbox{\boldmath $\nabla$}} \} \psi_i({\bf r}),
\label{kinetic-energy-density}
\end{equation}
and the spherical densities (\ref{densities}) and  therefore
the mesonic fields (\ref{mesons}) are all expressed in
terms of the Dirac spinors $\{\psi_i\}$.  In the RMF
approach Fock terms in Eq.(\ref{energy-density}) are
neglected.

Taking the variation of Eq.(\ref{energy-functional}) with
respect to $\psi_i^\dagger$ one  obtains the stationary
Dirac equation with the single-particle energies as
eigenvalues,

\begin{equation}
\hat h_D \psi_i({\bf r})~=~\varepsilon_i \psi_i({\bf r}),
\label{dirac}
\end{equation}
where

\begin{equation}
\begin{array}{ll}
\hat h_D ~=~-i{\mbox{\boldmath $\alpha$}}{\mbox{\boldmath $\nabla$}}
            + \beta (M + g_\sigma \sigma(r))
+g_\omega\omega^0(r) +g_\rho\tau_3\rho^0(r)
            + e{{(1-\tau_3)} \over 2}A^0(r).
\end{array}
\label{dirac-operator}
\end{equation}
Solving this equation self-consistently (the mesonic fields
depend  on the baryon  solution according to
Eq.(\ref{mesons})) one obtains the nuclear ground state
$\Phi_0$ in terms of the solutions $\{\psi_i\}$.

\section{Relativistic Generator Coordinate Method}
\label{rgcm}

The GCM has been used extensively within the
non-relativistic approaches to obtain the ground-state and
excited states of nuclei\cite{rin80}. Using the Skyrme
forces, the GCM was applied to study the giant
resonances\cite{flo76}.  Recently, the GCM has also been
employed to investigate the effect of correlations on the
ground-state properties of nuclei\cite{bon91}.  Here we
present a relativistic extention of the generator
coordinate method (GCM), which is based upon a trial
A-particle wave function ansatz $\Psi_{GCM}$ written in the
form of a linear combination:

\begin{equation}
\Psi_{GCM}({\bf r}_1\dots{\bf r}_A) =
\int {\cal F}(q)\Phi ({\bf r}_1\dots{\bf r}_A;q)dq
\label{hw-wf}
\end{equation}
where the generating function $\Phi(q)\equiv\Phi
({\bf r}_1,\dots{\bf r}_A;q)$ is chosen to be a Slater determinant
$\Phi(q)$ built upon single-particle spinors $\psi_i({\bf r},q),
(i=1,2,...,A)$, depending on the generator coordinate $q$.
It is obvious that in this case the wave function  of the
system (\ref{hw-wf}), being a superposition of Slater
determinants $\Phi(q)$, goes beyond the limits of the RMF
approach. The so-called 'weight', or 'generator' function
${\cal F}(q)$ is determined after varying  with respect to
${\cal F}(q)$  the  energy of the system

\begin{equation}
E[{\cal F}] = {{<\Psi_{GCM}|\hat H|\Psi_{GCM}>}\over
                  {<\Psi_{GCM}|\Psi_{GCM}>}} .
\label{hw-e}
\end{equation}
This leads to the Hill-Wheeler integral equation for the
weight function:

\begin{equation}
\int \left [ {\cal H}(q,q^{\prime }) - E{\cal N}(q,q^{\prime })\right]
 {\cal F}(q^{\prime })dq^{\prime } = 0 ,
\label{hw-eq}
\end{equation}
where

\begin{equation}
{\cal H}(q,q^{\prime }) = <\Phi(q)|\hat{H}|\Phi(q^\prime)>
\label{hw-h}
\end{equation}
and

\begin{equation}
{\cal N}(q,q^{\prime }) = <\Phi(q)|\Phi(q^\prime)>
\label{hw-n}
\end{equation}
are the energy and the norm overlap kernels, respectively.

A straightforward calculation shows that with the
Hamiltonian $\hat H$ associated with our model Lagrangian
(\ref{lagrangian}) one obtains

\begin{equation}
{\cal H}(q,q^{\prime })~\equiv~<\Phi(q)|\hat H |\Phi(q^\prime)>
\label{hw-hh}
\end{equation}
where ${\cal N}(q,q^{\prime })$ is the overlap kernel
(\ref{hw-n}) and ${\cal H }(r;q,q^\prime)$ is the overlap
energy-density kernel:

\begin{equation}
\begin{array}{ll}
 {\cal H }(r;q,q^\prime)&=~\tau(r;{q,q^\prime}) + M \rho_s(r;{q,q^\prime})
 + {1\over 2} g_\sigma\rho_s(r;{q,q^\prime})\sigma(r;{q,q^\prime}) \\
 &+ {1\over 2}\{ {2\over 3} g_2\sigma^3(r;{q,q^\prime})
                + {1\over 2} g_3\sigma^4(r;{q,q^\prime}) \}
  + {1\over 2} g_\omega\rho_v(r;{q,q^\prime})\omega^0(r;{q,q^\prime})\\
 &+ {1\over 2} g_\rho\rho_3(r;{q,q^\prime})\rho^0(r;{q,q^\prime})
  + {1\over 2} e\rho_p(r;{q,q^\prime})A^0(r;{q,q^\prime}).
\end{array}
\label{hw-hhd}
\end{equation}
In this equation the 'kinetic' energy density is defined by the spinors
$\{\psi_i({\bf r};q)\}$ as

\begin{equation}
\tau(r;q,q^\prime ) =\displaystyle{\sum_{i,j=1}^A}\; N^{-1}_{ji}\;
\psi_i^\dagger({\bf r};q) \{-i{\mbox{\boldmath $\alpha$}}
{\mbox{\boldmath $\nabla$}} \} \psi_j({\bf r};q^\prime ).
\end{equation}
Similarly the other densities entering Eq.(\ref{hw-hhd}) are:

\begin{equation}
\begin{array}{ll}
\rho_s(r;q,q^\prime )&=
\displaystyle{\sum_{i,j=1}^A}\; N^{-1}_{ji}\;
{\bar\psi}_i({\bf r};q) \psi_j({\bf r};q^\prime )\\
\rho_v(r;q,q^\prime )  &=\displaystyle{\sum_{i,j=1}^A}\; N^{-1}_{ji}\;
 \psi_i^\dagger({\bf r};q) \psi_j({\bf r};q^\prime )\\
\rho_3(r;q,q^\prime )  &=\displaystyle{\sum_{i,j=1}^A}\; N^{-1}_{ji}\;
 \psi_i^\dagger({\bf r};q)\tau_3\psi_j({\bf r};q^\prime )\\
\rho_p(r;q,q^\prime )
&=\displaystyle{\sum_{i,j=1}^A}\; N^{-1}_{ji}\;
 \psi_i^\dagger({\bf r};q){{(1-\tau_3)}\over 2}  \psi_j({\bf r};q^\prime ).
\end{array}
\label{hw-densities}
\end{equation}
They appear as source terms

\begin{equation}
    s_\phi(r;q,q^\prime ~=~\left\{~\begin{array}{ll}
-g_\sigma\rho_s(r;q,q^\prime )-g_2\sigma^2-g_3\sigma^3~~&\mbox{for
 the$\;\sigma$-field}\\
 g_\omega \rho_v(r;q,q^\prime )                       &\mbox{for
 the$\;\omega$-field}\\
 g_\rho \rho_3(r;q,q^\prime )                         &\mbox{for
 the$\;\rho$-field  }\\
 e\rho_p(r;q,q^\prime )                               &\mbox{for the Coulomb
 field,}
                        \end{array}
           \right.
\label{hw-sources}
\end{equation}
in Klein-Gordon equations of the type (\ref{kg}) whose
solution determines the fields
$\phi(r;q,q^\prime)$ = $\sigma(r;q,q^\prime )$, $\omega^0(r;q,q^\prime)$,
$\rho^0(r;q,q^\prime)$ and $A^0(r;q,q^\prime)$ entering
Eq.(\ref{hw-hhd}) as

\begin{equation}
\phi(r;q,q^\prime)=\displaystyle{\int_0^\infty}G_\phi(r,r')
                     s_\phi(r';q,q^\prime)r'^2 dr',
\label{hw-mesons}
\end{equation}
the Green functions are defined as before by Eqs. (\ref{green-mes}) and
(\ref{green-cou}).

In the above equations the sums run over all occupied
single-particle states and $ N^{-1}_{ij}$ are the elements
of the matrix $N^{-1}(q,q^\prime)$ where

\begin{equation}
N_{ij}(q,q^\prime)~=~
\int d^3 r\,\psi_i^\dagger({\bf r};q)\psi_j({\bf r};q^\prime ).
\end{equation}
The determinant of $N(q,q^\prime)$ simply gives the overlap
kernel (\ref{hw-n})

\begin{equation}
{\cal N}(q,q^{\prime }) =  \det\{N(q,q^\prime)\}.
\label{hw-nn}
\end{equation}

\par\noindent
Thus, having determined the integral kernels (\ref{hw-hh})
and (\ref{hw-nn}), the associated Hill-Wheeler integral
equation (\ref{hw-eq}) has to be solved in order to
determine the nuclear ground- and n-excited states through
its eigen solutions $\{E_0,{\cal F}_0(q)\}$ and
$\{E_n,{\cal F}_n(q)\}$, respectively.

\section{Isoscalar Giant Monopole Resonance}
\label{gmr-theory}

The constrained Hartree-Fock calculations have been a usual
method to obtain description of the excited states in
nuclei. An extension of this method in the framework of the
RMF theory has been made recently, where the breathing-mode
GMR in finite nuclei has been obtained in the constrained
calculations\cite{srs94}. We extend some of the discussion
here for the sake of clarity. In order to analyze the
isoscalar GMR we perform also constrained RMF calculations,
where the Dirac equation (see Eq.(\ref{dirac}))

\begin{equation}
\left(\hat h_D - q r^2\right) \psi_i(x) = \varepsilon_i \psi_i(x),
\label{c-dirac}
\end{equation}
is solved at different values of the Lagrange multiplier
$q$ which are associated with values of the nuclear $rms$
radius

\begin{equation}
R= \left\{{1\over A} \int r^2 \rho_v(r;q) d^3r \right\}^{1/2},
\label{rms}
\end{equation}
where

\begin{equation}
  \rho_v(r;q) = \displaystyle{\sum_{i=1}^A}\;
\psi_i^\dagger({\bf r};q) \psi_i({\bf r};q)
\end{equation}
is the baryon local density determined by the solution
$\{\psi_i({\bf r};q)\}$.  According to
Eq.(\ref{energy-functional}), the total energy of the
constrained system

\begin{equation}
E_{RMF}(q) = E_{RMF}[\psi_i(q)] ,
\label{c-energy}
\end{equation}
is a function of $q$ (or the nuclear $rms$ radius $R$). It
has a minimum, the ground-state energy
$E_{RMF}^0=E_{RMF}(0)$, at $q =0$ corresponding to the
ground-state $rms$ radius $R_0$. The curvature of this
function around the equilibrium point $R_0$ defines the
so-called constrained incompressibility coefficient of the
finite nucleus

\begin{equation}
K_C(A) =
 A^{-1} \left( R^2 {{d^2E_{RMF}(q)}\over {dR^2}} \right)_{q=0}.
\label{kc}
\end{equation}
The constrained energy (\ref{c-energy}) as a function of
$q$ represents the energy surface for the isoscalar
monopole motion of the nucleus where $R$ changes around its
ground-state value $R_0$.  In  order to derive vibrational
excitation energies one needs in addition the inertial
parameter for this motion.  In the non-relativisitic RPA
sum-rule approach (SRA) \cite{boh79} the inertia parameter
for the GMR is derived as $M R_0^2$.  In this case one
obtains the GMR excitation energy $E_1$ as
\begin{equation}
E_1 = \sqrt{ {K_C(A)\over { M R_0^2}}}.
\label{sra}
\end{equation}
In order to obtain a description of the GMR in the RMF
theory, we consider the Lagrange multiplier $q$ entering
Eq.(\ref{c-dirac}) as generator coordinate for the GCM
calculations as described in Sec.\ref{rgcm}. The solution
$\{\psi_i({\bf r};q)\}$ of Eq.(\ref{c-dirac}) at different
values $q$ then defines the generator Slater determinants
$\Phi(q)$ and therefore the integral kernels (\ref{hw-h})
and  (\ref{hw-n}). In fact, the diagonal part of the energy
kernel ${\cal H}(q,q^\prime )$ coincides with the
constrained energy (\ref{c-energy}), i.e.,
$E_{RMF}(q)\equiv {\cal H}(q,q)$.  The off-diagonal
elements ${\cal H}(q,q^\prime)$ contain the information
about the inertia. We then solve the resulting Hill-Wheeler
equation (\ref{hw-eq}) numerically using the method of
ref.\cite{flo76}.

\section{Results}
\label{results}

\subsection{Relativistic GCM Calculations}
\label{gcm-results}

We have performed GCM calculations for four closed-shell
nuclei $^{16}$O, $^{40}$Ca, $^{90}$Zr and $^{208}$Pb with
the sets of Lagrangian parameters given in
Table\,\ref{table1}.  The Lagrangian parameters sets are
NL1\cite{rei89}, NL-SH\cite{snr93}, NL2\cite{lee86},
HS\cite{hor81} and L1\cite{lee86} in the increasing order
of the nuclear matter incompressibility  with
$K_{NM}=\,$211.7, 355.0, 399.2, 545 and 626.3 MeV,
respectively. This allows us to examine the dependence of
GMR energies on the nuclear matter incompressibility
$K_{NM}$.  These sets of parameters have also been employed
in our earlier constrained RMF calculations\cite{srs94}.
The last two sets, HS and L1, correspond to the linear
model without the self-coupling of the $\sigma$-field.  In
addition, the set L1 excludes the contribution from the
$\rho$-field.  Among the sets NL1, NL-SH and NL2, which
correspond to the non-linear model, only the set NL2 has a
positive coupling constant $g_3$ in Eq (2). The set NL2 has
an effective mass $m^* = 0.67$ at the saturation, which is
higher than that of NL1 and NL-SH. Whereas the set NL1
reproduces the ground-state properties of nuclei only close
to the stability line due to the very large asymmetry
energy, the set NL-SH describes also nuclei far away from
the stability line\cite{snr93}. The shell effects at the
drip-line and deformation properties obtained with NL-SH
have been found to be in good agreement with the recent
finite-range droplet model\cite{mol94}.  In order to cover
the region of the incompressibility about 300 MeV, we have
also constructed a schematic force NL-S1 with $K_{NM} =
296$ MeV.  This force describes the ground-state properties
of closed-shell nuclei rather well, but has a large
asymmetry energy of 52 MeV.  The use of various parameter
sets in our GCM calculations allows to study the influence
of the model Lagrangian and their properties on the
properties of the GMR energies. The aim of the present
study is also to examine how the GCM works in the RMF
theory.

The self-consistent solution of the constrained mean field
problem (\ref{c-dirac}) diverges for large positive and
negative values of the constraining parameter $q$. For
large positive values the quadratic $r$-dependence of the
constraining operator leads to unbound solutions. For large
negative values of $q$, when the nuclear $rms$ radius $R$
decreases, the self-consistent mean-field potential pushes
up the single-particle energies and the RMF solution
disappears at some minimal $rms$ radius $R_{min}$. The
instability for negative $q$-values causes no further
problems. For positive $q$, however, we find only a very
limited range of possible $q$-values close to the ground
state. We therefore introduce a cut-off function
$(1+\exp(r-R_{cut}/a)^{-1})$ multiplying $r^2$ in the
constrained RMF calculations. The cut-off radius was chosen
to be $R_{cut}=2r_0A^{1/3}$ with $r_0 = 1.2$ fm. The
diffuseness parameter was set at $a = 0.5$ fm.  In Fig.
\ref{figure1} we show the first three excited states for
$^{90}$Zr obtained in the relativistic GCM calculations
with the force NL-SH.  The effect of the cut-off function
is demonstrated in this figure by two constrained RMF
calculations, one with (dot-dashed line) and the other
without (solid line) the cut-off function. The GCM kernel
without cutoff goes only up to 4.35 fm on the right side of
the ground-state as shown by the solid curve. It can be
seen that the inclusion of the cut-off function enlarges
the required space for the GCM calculations without
changing the behaviour of the integral kernels beyond the
one calculated without the cut-off function.

\subsection{The Correlated Ground State}
\label{ground-state}

Solving the Hill-Wheeler equation (\ref{hw-eq}) we obtain
the weight function $g_0= {\cal N}^{1/2} {\cal F}_0$\,
associated with the ground-state solution ${\cal F}_0(q)$.
Functions $g_0(q)$ show the usual bell shape with a maximum
around the RMF ground-state value $q$=0. With an increase
in mass number A the width of $g_0(q)$ decreases, while its
amplitude increases keeping fixed the normalization ${\cal
F}{\cal N}{\cal F} = 1$.

Typical GCM results for energies and $rms$ radii calculated
with the set NL1 are given in Table\,\ref{table2}. It is
worth noting that the GCM ground-state energy is slightly
lower than the RMF one. This small difference, which is too
small to be seen in Fig. 1 contains in fact two
contributions: (i) the positive zero-point energy of
roughly $1/2\hbar\omega$ in the harmonic approximation and
(ii) the correlation energy induced by the GCM-correlations
lowering the mean field energy of the ground state by
roughly the same amount.  This is an important point and
reflects the fact that GCM is beyond the RMF approximation.

There is also no perceptible effect on the $rms$ radii of
the nuclear ground state in the GCM.  The largest
difference between the RMF and the GCM ground-state $rms$
radii is seen for the nucleus $^{16}$O. It is about 0.0025
fm.  Fig.\,\ref{figure2} shows the RMF and the GCM vector
($\rho_v$) and scalar ($\rho_s$) densities. The RMF and GCM
local densities do not differ significantly. For heavy
nuclei the GCM ground-state local densities are even closer
to the uncorrelated RMF ones. We can thus conclude that the
correlations in the GCM ground-state are small and the main
purpose of the GCM consideration here is in its
possibility to generate nuclear excited states.

\subsection{GMR Excited  States}
\label{excited-states}

The first three GMR excited states obtained in the GCM with
the parameter set NL1 are shown in Table\,\ref{table2}.
These states show a clear equidistant spectrum for heavy
nuclei. A similar behaviour is also apparent from Fig. 1
for $^{90}$Zr too, which has been shown for the set NL-SH.
For the lighter nuclei, however, there are significant
deviations from this type of spectrum as can be seen from
the energies of the excited states in $^{40}$Ca and
$^{16}$O. With an increase in mass number, the excitation
energies decrease. Here we take the excitation energy
$\Delta E_1= E_1-E_0$, which is equivalent to the
excitation energy of a collective state in the
non-relativistic constrained Hartree-Fock approach. The
mass dependence $\Delta E_1= cA^{-1/3}$ of the excitation
energy for $^{208}$Pb, the nucleus on which there exists
well-measured GMR energy, is obtained as $c~=$ 69.1, 79.6,
93.6, 104.9, 97.0, and 126.2 MeV for the sets  NL1, NL-S1,
NL-SH, NL2, HS, and L1, respectively.

In Fig.\,\ref{figure2} we show the local vector and scalar
densities $\rho_{00}(r)$ of the GCM ground state and
$\rho_{11}(r)$ of the first excited GMR state for the set
NL1.  The densities are more extended in space in
comparison with the ground state ones. Consequently, the
$rms$ radii in the first exited state are larger than that
of the associated ground-state values by about 0.15 fm in
$^{16}$O and by only 0.015 fm in$^{208}$Pb.

\subsection{Transition Density}
\label{tran}

The transition density of the GMR provides the strongest
evidence for the radial density oscillations in nuclei and
hence of the 'breathing' or the compression character of
the GMR mode. We show in Fig.\,\ref{figure3} the vector and
scalar transition densities $\rho_{01}(r)$ for protons and
neutrons in $^{208}$Pb obtained in the relativistic GCM
calculations for the force NL-SH.  The transition densities
show a change in the density in the bulk at the expense of
that in the surface. A node at 6 fm is clearly to be seen
for protons and at about 6.4 fm for neutrons. The existence
of a well-defined node in the transition density is the
typical behaviour for the breathing-mode motion and
testifies for the compressional property of the GMR. The
transition density from RPA calculations for the GMR in
$^{208}$Pb with Skyrme force SIII was obtained to be very
similar to the transition density in Fig. 3. The transition
density for SIII also showed a node at about 6.2 fm. Both
these transition densities, one in the RMF theory for NL-SH
and the other in the Skyrme approach for SIII resemble much
that obtained from a simple radial scaling of ground-state
density. The point of difference to be noted is that in our
RMF case, we have obtained the transition density for the
GMR in the GCM, with some form of a constrained motion.
Here we do not observe any conspicuous differences between
the transition densities of the relativistic GCM and the
scaling mode in the Skyrme approach. The oscillations in
the interior of the nucleus are obviously due to the shell
effects. The vector transition density shown in the figure
conserves the particle number. The same can not, however,
be said for the scalar transition density, which is albeit
similar to the vector transition density, but manifests
mainly the relativistic effect similar to that exhibited by
total scalar density.

We observe that the difference between the scalar and the
vector transition densities, which is connected with the
small components of the Dirac wave functions, arises mainly
in the interior of the nucleus. In the surface region both
densities coincide more or less.

\subsection{Constrained Incompressibility of Finite Nuclei}
\label{compressibility}

We now consider the constrained incompressibility $K_C(A)$
as calculated from Eq.(\ref{kc}). The results as a function
of the nuclear matter incompressibility $K_{NM}$ are shown
in Fig.\ \ref{figure4} for a few nuclei. It may be worth
mentioning that empirically the GMR has been well
established only in heavy nuclei.  such as $^{208}$Pb and
$^{90}$Zr. We have also included the light nuclei such as
$^{40}$Ca, $^{16}$O and $^4$He. In the light nuclei it is
very uncertain and a full energy-weighted sum-rule strength
has rarely been observed. Thus, in our case the light
nuclei serve mostly the purpose of illustration and for the
possible anharmonic effects.

With exceptions for light nuclei, the incompressibility
$K_C(A)$ shows a strong dependence on the nuclear matter
incompressibility $K_{NM}$.  For the linear force HS,
$K_C(A)$ shows a slight dip from the increasing trend for
$^{208}$Pb and $^{90}$Zr, whereas for light nuclei
$^{40}$Ca, $^{16}$O and $^4$He, the HS values are even
smaller than the NL2 values.  The dependence of $K_C(A)$ in
the Skyrme approach is different, where it increases
monotonically with $K_{NM}$. For, the finite nuclear
incompressibility receives a sizeable contribution from the
surface incompressibility, this difference could be
explained from the difference in the behaviour of the
surface incompressibility in the two methods. In the Skyrme
approach, the surface incompressibility has been shown to
be $K_S \sim -K_{NM}$ for all standard Skyrme forces.  This
does not seem to be the case for the RMF theory, however,
as shown by the HS values. Thus, the surface
incompressibility is not necessarily a straight function of
the nuclear matter incompressibility in the RMF theory.
This point has also been dealt with in ref.\cite{srs94}.

\subsection{Comparison with Nonrelativistic Calculations
and Experimental Data}
\label{hf}

The excitation energy $\Delta E_1$ corresponds to the
energy $E^{(1)}$ usually obtained from the non-relativistic
constrained Skyrme Hartree-Fock (SHF) calculations within
the sum rule approach\cite{tre81}.  In Table\,\ref{table3}
energies  $\Delta E_1$ are compared with such
nonrelativistic constrained SHF results obtained with the
Skyrme-type forces SkM and SIII. These Skyrme forces have
nearly the same nuclear matter incompressibility $K_{NM}$
as do the sets NL1 and NL-SH, respectively.  It can be seen
that the nonrelativistic SHF results differ slightly from
the values of $\Delta E_1$. This difference in the
relativistic GCM excitation energy $\Delta E_1$ from the
SHF energy is small for heavy nuclei. It, however,
increases for lighter nuclei, where the GCM shows lower
values.

Fig.\ \ref{figure5} shows the GCM breathing-mode energy
$\Delta E_1$ for various nuclei and parameter sets. The
energy $\Delta E_1$ first increases with $K_{NM}$ from NL1
to NL2 almost linearly for all nuclei.  For the force HS,
which has $K_{NM}$ even larger than that of NL2, the energy
shows a decrease for all the nuclei, however. This is due
to a rather large surface incompressibility which is in
disproportion to its bulk incompressibility for HS. This
reduces the incompressibility of the nuclei, as also shown
in Fig. 4. For the force L1, $\Delta E_1$ shows an increase
compared to HS. Thus,  $\Delta E_1$ is not related in a
simple way to $K_{NM}$. This reflects the role played by
the surface component of the compression in the RMF theory.
Even for heavy nuclei  $\Delta E_1$ does not show an
overall increasing tendency with  $K_{NM}$.  For lighter
nuclei this effect is even more apparent.

It is interesting to note that the dip in energy  $\Delta
E_1$ for HS seems to signal the transition from nonlinear
(NL1, NL-SH, NL2) to linear (HS, L1) models in the
Lagrangian (\ref{lagrangian}). Even with significantly
higher nuclear matter incompressibility ($K_{NM}$= 545 MeV
for HS) the linear model gives comparable GMR excitation
energies (and even lower) in comparison with the nonlinear
ones (notice that $K_{NM}$= 399.2 MeV for NL2).

It is instructive to see that the approximate expression,
(\ref{sra}) which is exactly the same as in the
nonrelativistic sum-rule approach but calculated with the
incompressibility $K_C(A)$ emerging from the relativistic
RMF calculations, gives acceptable results for GMR
excitation energies. The results from Eq.(\ref{sra}) are
compared with the GCM and SHF results also in Table\
\ref{table3}.

In the non-relativistic approach using density-dependent
interactions, extensive work was carried out to obtain the
incompressibility and breathing-mode energy\cite{bla80}
with HF + RPA calculations. The RPA calculations were
performed on a set of Skyrme and Gogny interactions
including the finite-range Gogny force D1, with an
increasing order of incompressibility of nuclear matter.
This work attempted to reproduce the empirical
breathing-mode energies on $^{208}$Pb and $^{90}$Zr, where
experiments showed the existence of the GMR unambiguously.
The GMR energy in $^{208}$Pb is rather well-established and
lies at $13.7 \pm 0.3$ MeV. The GMR energy in $^{90}$Zr has
been measured to be in the range 16.5 - 17.3 MeV by
different experiments. The average value of the energy from
different experiment comes at about 17.0 MeV. Table
\ref{table4} shows the GMR values for $^{90}$Zr and
$^{208}$Pb obtained in the RPA calculations\cite{bla80} for
the forces D1, Ska and SIII. A comparison of the RPA values
with the empirical values in Table \ref{table4} shows that
the Gogny force D1 reproduces the GMR energy for $^{208}$Pb
quite well. The force D1, however, overestimates the GMR
energy for $^{90}$Zr by about 1.5 MeV. The force Ska, which
has incompressibility of nuclear matter at 263 MeV, gives
the GMR energy for $^{208}$Pb only slightly higher than D1.
The GMR energy with Ska for $^{90}$Zr is, however, about 2
MeV higher than the empirical value.  Thus, within the
non-relativistic approach, with D1 one comes very close to
reproducing the GMR energy of $^{208}$Pb in the RPA
calculations.  The GMR energy of $^{90}$Zr could not,
however, be reproduced by any Skyrme force. This has been
the scenario within the Skyrme approach, where the
conclusions of ref.\cite{bla80} on the incompressibility
hinged very strongly on $^{208}$Pb only.  Consequently, a
value of the incompressibility of nuclear matter of about
210 MeV seem to have been favoured.

We now compare the empirical values and the RPA results
with those in the relativistic GCM calculations with the
force NL-S1.  It may be noted that this force which
describes the ground-state properties of nuclei only very
approximately, and was constructed with a view to fill in
the gap at about $K_{NM} \sim 300$ MeV in the the
dependence of the incompressibility on the breathing-mode
energy. It has presently only a schematic character. With
its incompressibility of 296 MeV, the GMR energy for
$^{208}$Pb in the GCM has been obtained at 13.4 MeV. It is
very close to the empirical values obtained in many
experiments. The GCM energy for $^{90}$Zr has been obtained
at 17.6 MeV, which is slightly higher than the average
value of 17.0 MeV but is closer to an earlier empirical
result. On the whole, it is within the uncertainties of the
empirical values. In comparison, the GMR energy from the
Gogny force $D1$ in the RPA lies at 18.5 MeV. Systematics
of the values for $^{208}$Pb and $^{90}$Zr in Fig. 5 show
that both the empirical values as shown by the quadrangles
are encompassed by the GCM calculations curve from $K_{NM}
= 290-310$ MeV. The width of the quadrangles signify the
corresponding experimental uncertainties in the
determination of the GMR centroid energies. The empirical
values themselves have been reproduced by $K_{NM} \sim 300$
MeV as can be seen by intersecting the empirical values at
the curves for $^{208}$Pb and $^{90}$Zr.

\section{Conclusions and Discussion}
\label{conclusions}

We have performed a systematic study of the breathing-mode
energy and the incompressibility of finite nuclei with the
generator coordinate method in the RMF theory. It has been
observed that the transition density of the giant monopole
mode shows a character very similar to that obtained in the
Hartree-Fock-RPA approach with density-dependent Skyrme
interactions. This behaviour is also similar to what one
expects in the simple radial scaling of the ground-state
density.

Using a set of relativistic mean-field Lagrangian
parameters, it has been shown that the GCM energies for the
realistic forces show an increasing tendency with the
nuclear matter incompressibility. Only for unrealistic
forces such as HS does one observe a decrease in the
breathing-mode energy and also in the incompressibility of
nuclei even when this force has a larger $K_{NM}$. This is
due to a very large surface incompressibility of HS.

The GCM values obtained with the force NL1 are quite lower
than the empirical values and those with NL-SH are a little
higher than the latter. The empirical GMR energies, on the
other hand, can be well encompassed by the GCM curve from
$K = 280-310$ MeV. This is corroborated by the GCM values
obtained with a rather schematic force having an
incompressibility $K = 296$ MeV, where the GCM values for
$^{208}$Pb and $^{90}$Zr are very close to the
corresponding empirical values. Thus, the empirical GMR
energies for both these nuclei have been clearly bracketed
by the GCM calculations in the RMF theory. We know of no
other theoretical result where the GMR energies for both
these nuclei have been reproduced.  Our results also bring
about severe constraints on the value of the nuclear matter
incompressibility, the observable which has theoretically
been held rather uncertain. The GCM results, thus, favour
an incompressibility at about 300 MeV. This is in contrast
with the usual assumption of the incompressibility of about
210 MeV concluded from the non-relativistic Skyrme ansatz,
where the empirical value for $^{208}$Pb only could be
reproduced. Our conclusion, on the other hand, is in good
agreement with the analysis of the empirical breathing-mode
energies where the incompressibility of nuclear matter was
obtained as 300 MeV or higher\cite{sha89,sha91}. This
analysis is, however, not yet complete and further work on
it is in progress.

Differences in the shell-effects of the RMF theory and the
Skyrme approach and their implications on the ground-state
properties of nuclei such as isotope shifts\cite{slr93} and
on nuclei at drip lines\cite{slr94} have been discussed
earlier. The present work on the breathing-mode energies in
the GCM has brought about important differences also in the
dynamical properties of the RMF theory and the Skyrme
ansatz. The nearly good reproduction of the empirical GMR
energies in the relativistic GCM approach has become
possible due to the ratio of the surface incompressibility
to the bulk incompressibility, which has been obtained as
different from 1 in the RMF forces. This was also
demonstrated in ref.\cite{snr94} using various schematic
parameter sets that in the RMF theory it is possible to
obtain the ratio of the surface incompressibility to the
bulk incompressibility of up to about 2 or more. For the
realistic parameter sets NL1 and NL-SH, this ratio has been
shown\cite{srs94} to be higher than 1 (1.58 and 1.72
respectively) in the simple radial scaling of the
ground-state density in the semi-infinite nuclear matter
with the Thomas-Fermi approximation. In the Skyrme approach,
the ratio of 1 has essentially been at the origin of
problems in describing the mass dependence of the GMR
energies. Of course, there still remain some improvements
to be made in the ansatz of the RMF theory with a view to
describe accurately the ground-state energies of nuclei at
and far away from the stability line as is the case with
the force NL-SH as well as the dynamical properties such as
the breathing-mode energies in nuclei.

\acknowledgments
One of us (M.V.S.) would like to thank the kind hospitality
at TUM  M\"unchen.  This work is partly supported by the
EEC program {\it Go West} under the Contract No.
ERB-CHBI-CT93-0651 and the Bundesministerium f\"ur
Forschung und Technologie in Bonn under the project 06
TM733. We thank R. Behnsch for careful reading of the
manuscript.



\begin{table}
\caption{Parameter Sets for the Lagrangian (\protect
\ref{lagrangian}) }
\begin{tabular}{cddddd}
& NL1\cite{rei89}&NL-SH\cite{snr93}&
NL2\cite{lee86}&HS\cite{hor81}& L1\cite{lee86}\\
\tableline
$M$\tablenotemark[1] & 938.0 & 939.0 & 938.0 & 939.0 &938.0 \\
$m_\sigma$ & 492.25& 526.0592& 504.89& 520.0 & 550.0 \\
$m_\omega$ & 795.355& 783.0  & 780.0 & 783.0 & 783.0 \\
$m_\rho$   & 763.0 & 763.0   & 763.0 & 770.0 &   0.0 \\
$g_\sigma$ &10.138 & 10.44355& 9.111 & 10.47 & 10.30 \\
$g_\omega$ &13.285 &12.9451  &11.493 & 13.80 & 12.60 \\
$g_\rho$   & 4.975 &4.3828   & 5.507 & 8.07  &   0.0 \\
$g_2$      &$-$12.172&$-$ 6.9099&$-$2.304& 0.0 & 0.0 \\
$g_3$      &$-$36.265&$-$15.8337&  13.783& 0.0 & 0.0 \\
\hline
 & \multicolumn{5}{c}{Nuclear Matter Characteristics} \\
$M^*/M$            &  0.57 &   0.60 &   0.67&   0.54&   0.53\\
$K_{NM}$           & 211.7 &355.0 & 399.2 & 545.0 & 626.3 \\
$a_{sym}$          & 43.5  & 36.1  & 43.9  & 35.0  &      \\
\end{tabular}
\tablenotetext[1]{The masses, the incompressibility
$K_{NM}$ and the asymmetry energy $a_{sym}$ are in MeV, the
coupling $g_3$ in fm$^{-1}$.}
\label{table1}
\end{table}


\begin{table}
\caption{RMF and constrained GCM results for the
ground-state energies and mass $rms$ radii and for the
excitation energies $ \Delta E_n = (E_n-E_0)$ of the first
three monopole states calculated with the set NL1.}
\begin{tabular}{cddddddddd}
 & \multicolumn{2}{c}{ Energies (MeV)}&
  & \multicolumn{2}{c}{ Radii (fm)}&
  & \multicolumn{3}{c}{ Excitation Energies (MeV)} \\
\cline{2-3} \cline{5-6} \cline{8-10}
Nuclei&RMF&GCM& &RMF&GCM& &$\Delta E_1$&$\Delta E_2$&$\Delta E_3$ \\
\hline
$^{16}$O&-127.24&-127.46& &2.65&2.65& &20.6&38.9&49.8 \\
$^{40}$Ca&-342.48&-342.58& &3.38&3.38& &17.1&29.9&37.3 \\
$^{90}$Zr&-784.90&-784.99& &4.28&4.28& &14.7&29.1&43.1 \\
$^{208}$Pb&-1639.89&-1640.07& &5.67&5.67& &11.7&23.3&34.9 \\
\end{tabular}
\label{table2}
\end{table}


\begin{table}
\caption{Comparison of the GMR excitation energies (in MeV)
obtained within the constrained GCM calculations and the
approximation ({\protect \ref{sra}}) using the constrained
incompressibility ({\protect \ref{kc}}) with the
nonrelativistic sum-rule approach obtained within
nonrelativistic constrained Hartree-Fock (HF) calculations
{\protect \cite{boh79}}. In the relativistic case sets NL1
and NL-SH are used, which have nearly the same nuclear
matter incompressibility $K_{NM}$ as do the sets of Skyrme
force parameters SkM and SIII used in the nonrelativistic
HF calculations {\protect \cite{tre81}}. }
\begin{tabular}{cccccccc}
 & \multicolumn{3}{c}{{NL1:}  {$K_{NM}=211.7$\,MeV}} &
 & \multicolumn{3}{c}{{NL-SH:} {$K_{NM}=354.95$\,MeV}} \\
 & \multicolumn{3}{c}{{SkM:} {$K_{NM}=216.7$\,MeV}} &
 & \multicolumn{3}{c}{{\ \ \ \ SIII:} {$K_{NM}=356.00$\,MeV}} \cr
\cline{2-4} \cline{6-8}
Nuclei&GCM &Eq.(\ref{sra})&Skyrme HF&
      &GCM &Eq.(\ref{sra})&Skyrme HF  \\
\hline
$^{16}$O  &20.6&20.9&22.4& &25.3&25.8&26.6 \\
$^{40}$Ca &17.1&19.2&20.2& &22.4&23.9&24.7  \\
$^{90}$Zr &14.7&16.3&17.0& &20.16&21.1&21.2 \\
$^{208}$Pb&11.7&12.2&12.9& &15.8&16.1&16.2
\end{tabular}
\label{table3}
\end{table}


\begin{table}
\caption{Comparison of the RMF results from the parameter
set NL-S1, with the HF+RPA calculations using
density-dependent Skyrme interactions.  Here we show the
results only for the nuclei $^{208}$Pb and $^{90}$Zr, where
the empirical data is reliable and rather
well-established.}

\begin{tabular}{cccccccc}
 & \multicolumn{3}{c}{Skyrme interactions+RPA} &
 & \multicolumn{1}{c}{GCM-RMF} \\
 & \multicolumn{1}{c}{ D1} & {Ska} & {SIII}
 & \multicolumn{2}{c}{~~~NL-S1}
 & \multicolumn{2}{c}{expt.} \\
 $K_{NM}$  & \multicolumn{1}{c}{228} & {263} & {356}
 & \multicolumn{2}{c}{296}\cr
\hline
$^{90}$Zr & 18.5 & 19.1 & 22.1 && 17.6 && $17.0 \pm 0.5$  \\
$^{208}$Pb & 14.4 & 14.7 & 17.2 && 13.4 && $13.5 \pm 0.3$ \\
\end{tabular}
\label{table4}
\end{table}

\begin{figure}
\caption{The constrained energy Eq.(\protect\ref{c-energy})
as a function of the $rms$ radius $R$ in Eq.
(\protect\ref{rms}), for $^{90}$Zr with the force NL-SH in
the GCM calculations.  The solid curve on the left is
without taking the cut-off function into account. For the
lower densities (higher $R$) the constrained energy has
been extended (dot-dashed line) by inclusion of the cut-off
function. The harmonic approximation to the curve is shown
by the dotted parabola. The energies of the first three
excited states ($E_1$, $E_2$ and $E_3$) along with the
ground-state energy $E_0$ are also shown. The energy $E_1$
in the harmonic approximation (dotted line) differs only
slightly from the actual value.}
\label{figure1}
\end{figure}

\begin{figure}
\caption{The ground-state vector and scalar densities for
protons in the RMF ($\rho_{RMF}$) and in the GCM
($\rho_{00}$) for $^{16}$O.  The effect of the ground-state
correlations in the GCM densities are seen to be minimal.
The GCM density for the first excited state of the GMR
($\rho_{11}$) is also shown, where there is a considerable
change in the densities in the interior. A slight change in
the surface of the nucleus can also be seen.}
\label{figure2}
\end{figure}

\begin{figure}
\caption{The vector and the scalar transition density for
the GMR in $^{208}$Pb obtained in the relativistic GCM
calculations with the force NL-SH. There is a conspicuous
node in the densities of both the protons and neutrons. The
change in the bulk of the vector density takes place at the
expense of the change in the surface, thus conserving the
total number of particles.}
\label{figure3}
\end{figure}

\begin{figure}
\caption{The constrained incompressibility $K_C$ obtained
in the RMF theory using various parameter sets. $K_C$
increases monotonically from NL1 to NL-SH for all nuclei.
The values for HS show a slight dip, indicating a very
large surface incompressibility.}
\label{figure4}
\end{figure}

\begin{figure}
\caption{The energy $\Delta E_1$ of the GMR obtained with various
relativistic Lagrangian sets in the GCM. The empirical values
of the GMR in $^{208}$Pb and $^{90}$Zr have been shown at their
average values by horizontal quadrangles. The widths of the
quadrangles span the error bars in the empirical data.
The empirical data encompass the corresponding GCM results at about
$K = 280 - 310$ MeV and show a good agreement with the
values obtained with the set NL-S1.}
\label{figure5}
\end{figure}
\end{document}